# New Phenomena Beyond Both the Standard Model and MSSM [1] [2]

JoAnne L. Hewett

*Stanford Linear Accelerator Center, Stanford University*
*Stanford, CA 94309, USA*

Signals for new, non-supersymmetric physics at hadron colliders are reviewed. We focus on extended gauge sectors and new matter particles.

## OVERVIEW

The Standard Model (SM) is in complete agreement with present experimental data. Nevertheless, it is believed to leave many questions unanswered, and this belief has resulted in numerous attempts to find a more fundamental underlying theory. One key ingredient in the extrapolation of the SM to higher energies is to identify the complete particle spectrum at the electroweak scale. Two popular examples of theories which populate the TeV scale with a plethora of new particles are supersymmetry and technicolor. This has resulted in extensive searches for super- and techni-particles, which have been reported elsewhere at this meeting (1). In this talk, I will identify other possible manifestations of new physics, and discuss their implications on hadron collider physics.

## EXTENDED GAUGE SECTORS

The phenomenology of models with extended gauge symmetries is particularly rich with the existence of new gauge bosons being the hallmark signature of such theories (2). However, additional gauge bosons are not the sole manifestation of an extended gauge group, as these theories also contain exotic fermions, which are required for anomaly cancellation, as well as an enlarged Higgs sector to facilitate the extended symmetry breaking. In addition, Supersymmetry may also be present, particularly in Grand Unified Theories

---

[2]Presented at the *10th Topical Workshop on Proton-Antiproton Collider Physics*, Batavia, IL, 9-13 May, 1995
[1]Work Supported by the Department of Energy, Contract DE-AC03-76SF00515

(GUTS), in order to solve the hierarchy problem and to ensure coupling constant unification.

Perhaps the most appealing set of enlarged electroweak models are those which are based on SUSY-GUTS, examples being the unifying groups SO(10) and $E_6$. In $E_6$ effective rank-5 models (3), additional neutral gauge bosons arise from the symmetry breaking chain

$$E_6 \to SO(10) \times U(1)_\psi \to SU(5) \times U(1)_\chi \times U(1)_\psi$$
$$\to SM \times U(1)_\theta, \qquad (1)$$

where $U(1)_\theta$ is a linear combination of $U(1)_{\chi,\psi}$ and remains unbroken at low energies ($\lesssim 1$ TeV). The parameter $\theta$ governs the fermion couplings of the $Z'$ boson and lies in the range $-90° \leq \theta \leq 90°$. Special models of this type include $\theta = 0°$ (Model $\psi$), $\theta = -90°$ (Model $\chi$) and $\theta = \arcsin(\sqrt{3/8})$ (Model $\eta$). We note that in the GUTS Renormalization Group Evolution of this model, the extra $U(1)$ enters at the 2-loop level and that there are additional low mass thresholds from the new particle content. $E_6$ and SO(10) GUTS can also lead to the symmetry chain

$$SO(10) \to SU(3)_c \times SU(2)_L \times SU(2)_R \times U(1)_{B-L}, \qquad (2)$$

which yields right-handed charged currents (as well as an additional neutral current) and is the now classic (4) left-right symmetric model (LRM). In this model $\kappa \equiv g_R/g_L$ represents the ratio of the right- to left-handed current coupling strengths, and lies in the range (4) $0.55 \lesssim \kappa \lesssim 2.0$. It has been shown (5) that a light right-handed mass scale ($\sim 1$ TeV) is consistent with coupling constant unification in Supersymmetric SO(10) models. We note the existence of a right-handed Cabbibo-Kobayashi-Maskawa (CKM) matrix in this model, $V_R$, which need not be the same as the corresponding left-handed mixing matrix. Another extended model based on the above 'low-energy' gauge group is the alternative left-right symmetric model (ALMR) (6), which is embedded in $E_6$ GUTS and switches the quantum number assignments between some of the ordinary and exotic fermions contained in the **27** representation of $E_6$. In this case, $\kappa = 1$, and the right-handed $W$ carries lepton number, has negative R-parity, and is produced via the parton-level reaction $gu \to W_R+$leptoquark, thus avoiding the usual mass constraints on right-handed $W$'s.

There is also a large number of extended electroweak models which are not based on a GUTS scenario. The principal case of this type is that of the sequential SM (SSM), in which the additional $Z$ boson is an exact replica of the SM $Z$, only heavier. This model is not gauge invariant, but it provides a useful benchmark in judging the search capabilities of various experiments. A list of other models in this category can be found in Refs. (2,7).

In all of the above extended electroweak models, the $Z - Z'$ mass matrix takes the form

$$\mathcal{M}^2 = \begin{pmatrix} M_Z^2 & \gamma M_Z^2 \\ \gamma M_Z^2 & M_{Z'}^2 \end{pmatrix}, \qquad (3)$$

**TABLE 1.** Mass bounds (in GeV) on new neutral gauge bosons from a fit to precision electroweak data performed in Ref. 8.

| Model | Unconstrained Fit | Constrained Fit |
| --- | --- | --- |
| $\chi$ | 330 | 920 |
| $\psi$ | 170 | 170 |
| $\eta$ | 220 | 610 |
| LRM | 390 | 1360 |
| SSM | 960 | |

where $\gamma$ is determined in each model once the Higgs sector is specified. The physical eigenstates are then

$$Z_1 = Z' \sin\phi + Z \cos\phi ,$$
$$Z_2 = Z' \cos\phi - Z \sin\phi , \qquad (4)$$

where $Z_1$ is presently being probed at LEP, and $\tan 2\phi = 2\gamma M_Z^2/(M_Z^2 - M_{Z'}^2)$, with the constraint $|\phi| \lesssim 0.01$ from LEP data (8). Restrictions on extended gauge sectors can be obtained from precision measurements by limiting the $Z_2$ contributions to processes such as $\mu$-decay, deep-inelastic neutrino scattering, atomic parity violation, as well as from the properties of the $Z_1$ boson, and the mass of the $W$ boson. These indirect bounds are summarized (2) in Table 1, from the results of a global electroweak fit performed in Ref. (9). The limits are presented for (i) an unconstrained fit with no assumptions on the Higgs sector, and (ii) a constrained fit where $\gamma$, and hence $\phi$ are specified for a given Higgs sector. The shift in the $W$ mass due to $Z-Z'$ mixing is presented in Fig. 1 from Ref. (10), where we see that a measurement of $\delta M_W \sim 100$ MeV would provide stringent bounds on $|\phi|$. The dashed line in this figure represents the constraint $\delta M_W = M_Z \beta \phi \sqrt{1-x_w}/2$ with $\beta < 1.5$ for all models discussed here.

New gauge bosons can be produced directly at hadron colliders via (i) the Drell-Yan mechanism, $p(\bar{p}) \to Z' \to \ell^+\ell^-$ and $p(\bar{p}) \to W'^{\pm} \to \ell^{\pm} \not{p}_T$, for which the new gauge boson must couple to both quarks and leptons, and (ii) the $Z', W' \to$ 2-jets channel. The latter mechanism requires the observation of a peak in the inclusive dijet invariant mass spectrum and is discussed at this meeting by Harris (11). The 95% C.L. $Z'$ search reach in the Drell-Yan channel at the Tevatron is displayed in Fig. 2 (a) for various models as a function of integrated luminosity, (b) for the rank-5 $E_6$ model as a function of $\theta$, and (c) for the LRM as a function of $\kappa$. These results are for electron data samples alone; the inclusion of muon final states would increase the mass reach by $\simeq 35 - 40$ GeV. In all cases we assume that the $Z'$ decays only to SM fermions, $Z - Z'$ mixing is neglected, and the CTEQ2M parton densities are used. We see that $Z'$ masses up to 1 TeV will be probed with the main injector luminosity upgrade. The new CDF search limit reported at this meeting (12) of $M_{Z'} > 650$ GeV for the SSM with $\mathcal{L} = 70 pb^{-1}$ in the electron

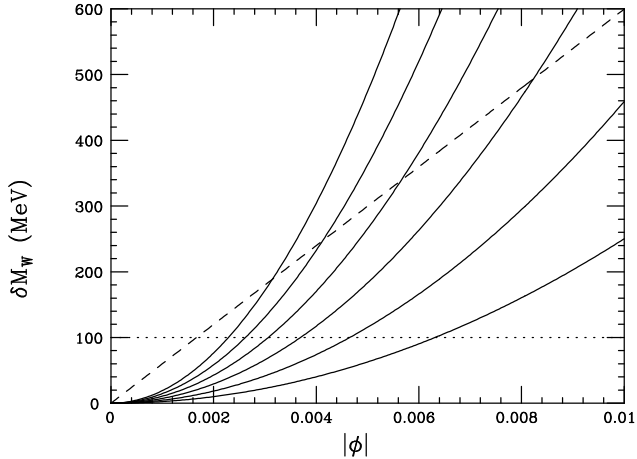

**FIG. 1.** $W$ mass shift due to $Z - Z'$ mixing. From top to bottom the curves correspond to $Z'$ masses of 2.0, 1.75, 1.5, 1.25, 1.0, 0.75 TeV.

+ muon channel agrees well with our expectations. The $Z'$ search capability of a $\sqrt{s} = 14$ TeV LHC is presented in Fig. 2d as a function of integrated luminosity, with the same set of assumptions as listed above. These search limits would degrade if exotic decay channels were open to the $Z'$ boson. For example, the LHC search reach is reduced by $\simeq 300 - 400$ GeV if the $Z'$ leptonic branching fraction decreased by a factor of 2.

*If* a new neutral gauge boson is discovered, a much more interesting and difficult question arises, *e.g.*, from which extended electroweak gauge model does the new $Z'$ originate? Numerous studies of this issue have been performed and are summarized in Ref. (2). Several processes have been proposed as a means to determine the couplings of the $Z'$ to the SM quarks and leptons, including the (i) leptonic forward-backward asymmetry, (ii) $\tau$ polarization asymmetry, (iii) 3-body $Z'$ decays $Z' \rightarrow \ell\nu W, \nu\bar{\nu}Z$, (iv) associated production $p(\bar{p}) \rightarrow Z'\gamma, g, Z, W$, (v) rapidity ratios, and (vi) examining the 2-jet decay of the $Z'$. All of these techniques (except for the forward-backward asymmetry) suffer from either (or both) a large background, and an event rate which dies off at $M_{Z'} \gtrsim 1-2$ TeV at the LHC. Naive parton-level studies have been carried out (2) and show that the $Z'$ couplings can be determined at the $\sim 5-20\%$ level with $100 fb^{-1}$ at the LHC for $M_{Z'} = 1$ TeV.

The corresponding search for an additional charged gauge boson via the Drell-Yan mechanism relies on the assumptions that (i) the $W'$ production vertex has SM coupling strength, and (ii) the $W'$ decays into a light and stable neutrino which manifests itself in the detector as missing $E_T$. Here we explore the ramifications of each assumption in the context of the LRM. In the LRM, the first supposition corresponds to taking $\kappa = 1$ and to setting the right-handed CKM matrix equal to its left-handed counterpart. The $W_R$ search reach as a function of $\kappa$ (taking $V_R = V_L$) is presented in Figs. 3(a-b) for various integrated luminosities at the Tevatron and LHC, respectively. Again, we see that masses of order 1 TeV will be explored at the Tevatron with the main injector, and that the LHC can search for masses up to the $4-5$ TeV range, with a decrease of $\sim 500$ GeV if the leptonic branching fraction is reduced by a factor of 2. A more interesting variation in the $W_R$ search reach results if the assumption $V_R = V_L$ is relaxed. A Monte Carlo study performed by Rizzo (13) shows that a significant search reach degradation can occur at the Tevatron in this case. Fig. 3(c) shows the percentage of $V_R$ parameter space which allows a $W_R$ below a given mass for Run Ia. We see that for $50(10)\%$ of the parameter space, the Run Ia search limit is reduced to $M_{W_R} \gtrsim 550(400)$ GeV (the bound assuming (i) above is 652 GeV (12)). This reduction is the result of modifying the weight of the various parton densities which enter the production cross section. The corresponding spread in the cross section $\times$ leptonic branching fraction at the LHC is displayed in Fig. 3(d). At the LHC, surrendering the hypothesis $V_R = V_L$ is not as costly since the Drell-Yan $W_R$ production occurs through sea $\times$ valence parton densities in $pp$ collisions, whereas the process is nominally valence $\times$ valence at the Tevatron if $V_{R_{ud}} \simeq 1$. The $W_R$ search becomes more problematic if assumption (ii) above is surrendered and the right-handed neutrino is massive. If the $W_R$ decay into the right-handed neutrino is kinematically allowed, one then has events of the type $W_R^+ \to \ell^+ \nu_R \to \ell^+ \ell^\pm + jj$, where either lepton charge sign is equally likely if $\nu_R$ is a Majorana fermion. Searches of this type were reported at this meeting by D0 (14) with the general result $M_{W_R} \gtrsim 520$ GeV for some regions of the parameter space. The worst case scenario results when $m_{\nu_R} > M_{W_R}$ so that the $W_R$ has only the hadronic (or exotic) decay channels open. In this case, one must search for bumps in the dijet distributions (11). Additional help may be gained by making use of the $W_R - Z_R$ mass relationship in the LRM, *i.e.*, if a $Z_R$ is found then this relationship tells us something about where to look for the right-handed $W$ in the dijet channel. If instead, only a limit on the $Z_R$ mass is obtained, then this mass relationship can be used and yields (10) a relatively weak bound on the $W_R$.

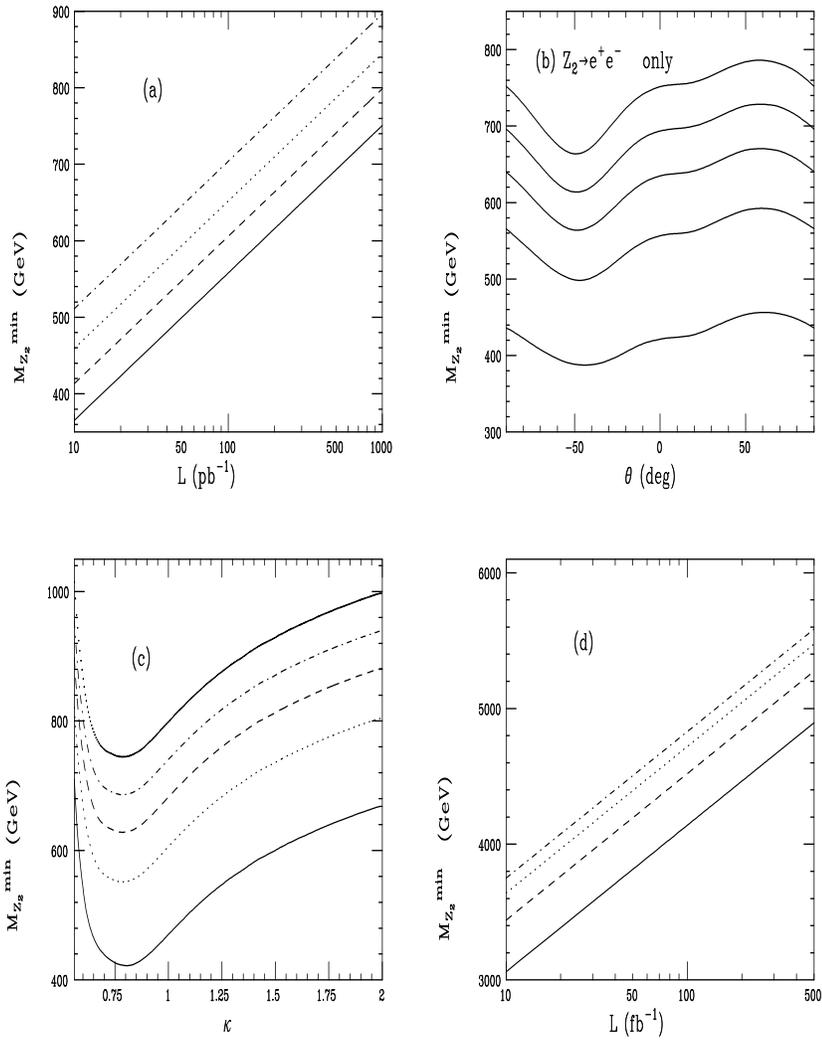

FIG. 2. $Z'$ search reach at the Tevatron (a) as a function of integrated luminosity for model $\psi$ (solid), the LRM (dashed), the SSM (dotted), and the ALRM (dash-dotted); (b) for the rank-5 $E_6$ model as a function of the parameter $\theta$, for the integrated luminosities of Run Ia, 100, 250, 500, 1000 $pb^{-1}$, from bottom to top; (c) for the LRM as a function of the ratio of right- to left-handed coupling strengths with the same values of integrated luminosity as in (b). (d) Same as in (a), except for the LHC.

## NEW MATTER PARTICLES

Excluding supersymmetric particles, new matter particles may be classified by three categories (15): exotic fermions, excited fermions, and difermions. Exotic fermions are predicted by many gauge extensions of the SM and include sequential (fourth generation) fermions, vector fermions (*e.g.*, those present in $E_6$ GUTS), mirror fermions, and singlet fermions. If a new gauge boson associated with an extended gauge group is found to be relatively light, then unitarity arguments force (16) the exotic fermions in the theory to also be light. All color triplet exotic fermions are copiously produced at hadron colliders via standard QCD processes, with signatures depending on the decay kinematics of the particular model. For example, one 'gold-plated' signal (17) for the iso-scalar quark present in $E_6$ theories is $p(\bar{p}) \to Q\bar{Q} \to (qZ)(\bar{q}Z) \to (j\ell^+\ell^-)(j\ell^+\ell^-)$, which occurs at large rates at the LHC for $m_Q \lesssim 1$ TeV. Indirect signals for very heavy quarks may also be potentially observable at the LHC (18) from the triangle and box diagram contributions to $gg \to ZZ$. Heavy lepton production can also be important in hadronic collisions, and proceeds through the Drell-Yan mechanism, gluon fusion (via the triangle diagram), and photon-photon fusion. The resulting cross sections (15) are $\sim 10^2$ fb at the LHC for heavy lepton masses in the few hundred GeV range.

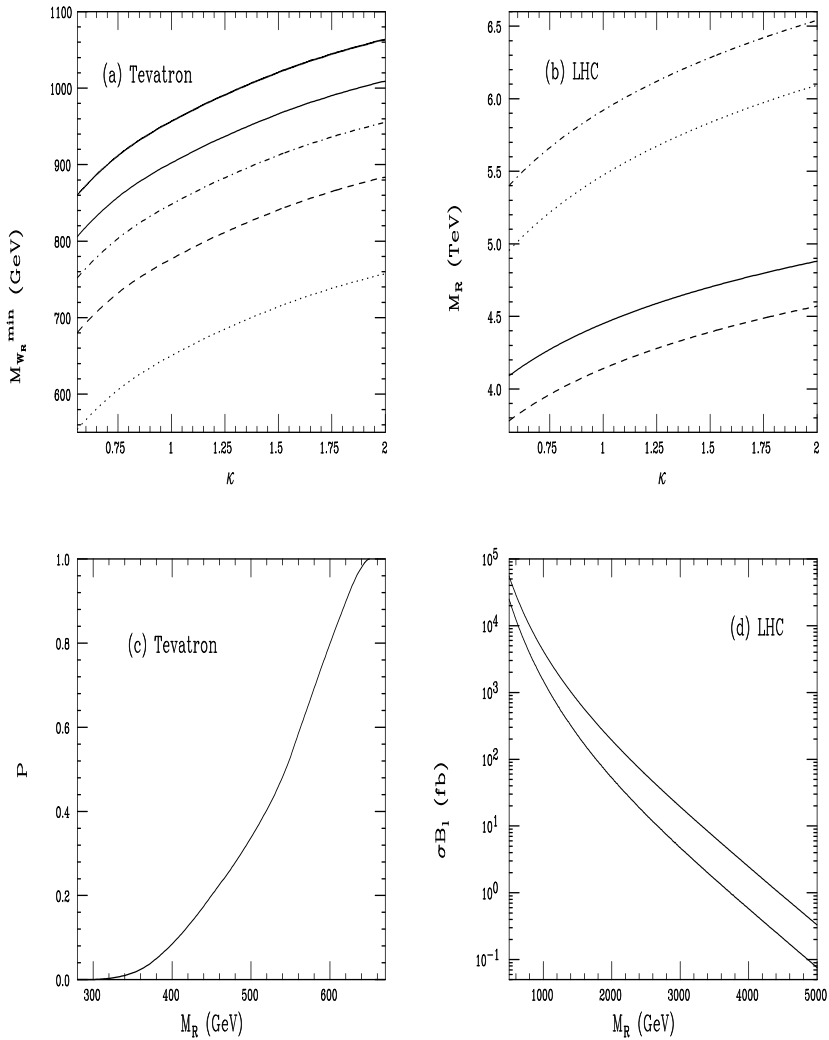

**FIG. 3.** $W_R$ search reach as a function of the ratio of right- to left-handed coupling strengths for (a) the Tevatron with the same values of integrated luminosity as in Fig. 2(b); (b) the LHC with 100 $fb^{-1}$, with the top (bottom) 2 curves corresponding to $sqrts = 14(10)$ TeV. In each set of curves, the bottom curve represents the reduction in search capability when the leptonic branching fraction is decreased by a factor of 2. (c) Percentage of the $V_R$ parameter space allowing the $W_R$ mass to be below a given value from Run Ia. (d) Maximum and minimum cross sections for $W_R$ production at the LHC as allowed by the $V_R$ parameter space (taking $\kappa = 1$).

The second category of new non-SUSY matter particles, *i.e.*, excited fermions, is a characteristic signature for substructure in the fermionic sector. Compositeness is a potential alternative to the SM description of electroweak symmetry breaking and it is conceivable that the first excitations will not make their presence felt until the Fermi scale, or above, is explored. Excited quarks can be produced in hadron collisions via the parton level processes $g + q \to q^*$ and $qq \to qq^*, q^*q^*$, with the characteristic signatures of dijet mass bumps, or jet + gauge boson or jet + lepton pair combinations. The existence of excited leptons can also be probed via contact interactions, where the cross sections could be large with distinctive leptonic final states. More details on exotic fermion production at hadron colliders can be found in, *e.g.*, Ref. (11,15).

Difermions can be either scalar or vector particles with unusual baryon or lepton number assignments, such as diquarks, dileptons, and leptoquarks. Diquarks are, of course, copiously produced via s-channel resonance at hadron colliders and yield bumps in the dijet mass spectrum (3,11). Leptoquarks are naturally present in theories which place quarks and leptons on an equal footing, such as SU(5), SO(10), and $E_6$ GUTS, technicolor, and composite models. They couple to a lepton-quark pair via a Yukawa type coupling of unknown strength; this is often parameterized as $\lambda^2/4\pi = F\alpha_{em}$. They also have the usual gauge couplings to the photon, the $Z$ and $W$ bosons, and gluons (for which an anomalous magnetic moment can exist in the case of vector leptoquarks). A systematic classification of the possible leptoquark quantum number assignments can be found in Buchmüller *et al.* (19), which yields 10 types of leptoquarks, 5 of which are scalar, and 5 being vector particles. Leptoquarks may be pair produced in hadronic collisions via $gg, qq \to LQ\,\overline{LQ}$, similar to squark pair production. These processes are essentially independent of the unknown Yukawa coupling $\lambda$, and yield the signatures 2jets $+\ell^+\ell^-$, $+\ell^\pm\,\not{p}_T$, or $+\,\not{p}_T$. The total production cross sections at the Tevatron for scalar (20) and vector (21) leptoquark pair production are given in Figs. 4(a) as a function of the leptoquark mass, and in (b) as a function of the possible anomalous magnetic moment $\kappa$ in the vector case. The on-going searches for the pair production of these particles are summarized by Park (12) and Lueking (22) at this meeting. Leptoquarks may also be singly produced (20) through the mechanism $gq \to LQ + \ell, LQ + \nu$, with the total cross sections being presented in Figs. 4(c-d) for scalar and vector leptoquarks, respectively (taking $F = 1$). The possible signatures for single production, jet $+\ell^+\ell^-$, $+\ell^\pm \not{p}_T$, or $+ \not{p}_T$, are distinctive. This process has the disadvantage in that it is directly proportional to the unknown value of $\lambda^2$, but it does have a large amount of available phase space, and hence yields larger rates than that from pair production for the parameter values $F \gtrsim 0.1$. The search reach for this process, *albeit* dependent on $\lambda$, could extend the bound obtained from pair production, and could be competitive with that of HERA. Another single production process (23) is given by photon bremsstrahlung from an initial quark and yields manageable event rates at the LHC. Indirect signatures for leptoquark exchange (24) could affect the Drell-Yan distribution $q\bar{q} \to e^+e^-$. This t-channel leptoquark exchange goes as $\lambda^4$, and hence produces an observable effect only for very large values of the parameter $F$.

## CONCLUSIONS

In summary, we see that signatures for new physics at hadron colliders are many and diverse, and we urge our experimental colleagues to continue to hunt for them.

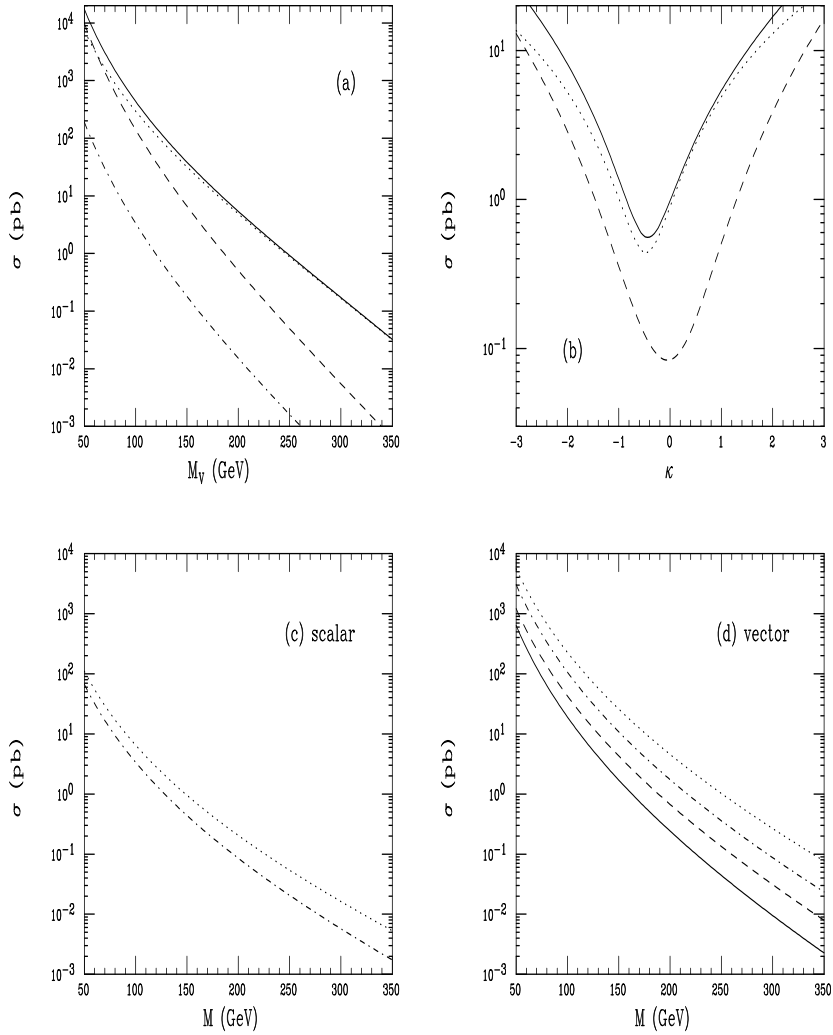

**FIG. 4.** (a) Cross section at the Tevatron for scalar leptoquark pair production (dash-dotted); the gg (dashed), $q\bar{q}$ (dotted), and total (solid) contributions for vector leptoquark pair production. (b) Cross section for vector leptoquark pair production at the Tevatron as a function of the anomalous magnetic moment for the gg (dashed), $q\bar{q}$ (dotted) and total (solid) contributions. (c-d) Single leptoquark production at the Tevatron with $F = 1$ for (c) scalars where the dotted (dash-dotted) curve represents the gu (gd) parton contributions, (d) vectors where the dotted (dash-dotted) curves represent the gu parton contributions with $\kappa = 1(0)$ and the dashed (solid) curves correspond to the gd parton contributions with $\kappa = 1(0)$.